\documentclass{ws-ijmpcs}

\usepackage{amssymb, amsopn, amsmath} 
\def\trimmarks{}

\begin{document}

\markboth{Chachamis {\it et. al.}}
{High energy factorization at NLO}

%
\catchline{}{}{}{}{}
%

\title{HIGH ENERGY FACTORIZATION AT NLO: LIPATOV'S EFFECTIVE ACTION REVISITED}

\author{G. CHACHAMIS}

\address{ Instituto de F\'isica Corpuscular UVEG/CSIC,
E-46980 Paterna (Valencia), Spain  \\
grigorios.chachamis@ific.uv.es}

\author{M. HENTSCHINSKI\footnote{Speaker}}

\address{Physics Department, Brookhaven National Laboratory, Upton, NY 11973,
USA. \\ hentsch@bnl.gov}

\author{J. D. MADRIGAL MART\'INEZ}
\address{Instituto de F\'isica Te\'orica UAM/CSIC, Nicol\'as Cabrera 15 \& \\ 
 Universidad Aut\'onoma de Madrid, C.U. Cantoblanco, E-28049 Madrid, Spain.\\
josedaniel.madrigal@uam.es}

\author{A. SABIO VERA}

\address{Instituto de F\'isica Te\'orica UAM/CSIC, Nicol\'as Cabrera 15 \& \\ 
 Universidad Aut\'onoma de Madrid, C.U. Cantoblanco, E-28049 Madrid, Spain.\\
agustin.sabio@uam.es}
\maketitle

\begin{history}
\end{history}

\begin{abstract}
We discuss aspects of our recent derivation of the gluon Regge trajectory at two loop from Lipatov's high energy effective action.  We show how the gluon Regge trajectory can be rigorously defined through renormalization of the high energy divergence of the reggeized gluon propagator. We furthermore provide some details on the determination of the two-loop reggeized gluon self-energy.

\end{abstract}

\ccode{PACS numbers: 11.25.Hf, 123.1K}

\section{Introduction}\label{1}

Balitsky-Fadin-Kuraev-Lipatov (BFKL) resummation\cite{BFKL1,BFKLNLO}
and high energy factorization provide the basis for the study of the
high energy limit of hard QCD scattering processes.  Recent
phenomenological evidence for BFKL evolution is found in the analysis
of the combined HERA data on the structure function $F_2$ and
$F_L$\cite{Ellis:2008yp,Hentschinski:2012kr} and the study of
di-hadron spectra in high multiplicity distributions at the Large
Hadron Collider\cite{Dusling:2012cg}.  An interesting application to
phenomenology is furthermore provided by Transverse-Momentum-Dependent parton distribution
functions, especially in region of small parton momentum fraction  $x$, where high energy factorization provides a natural definition,  see
Ref.~[\refcite{upd}] for recent work.\\

An attractive approach for the theoretical investigation of high
energy factorization is provided by Lipatov's gauge invariant high
energy effective action\cite{LevSeff} which provides a re-formulation
of QCD at high center of mass energies as an effective field theory of
reggeized gluons.  In Ref.~[\refcite{quarkjet}] a scheme has been
developed which comprises regularization, subtraction and
renormalization of high energy divergences and allows to use the high
energy effective action for the determination of next-to-leading order
corrections to high energy factorized matrix elements. So far this
scheme has been successfully applied to the derivation of forward jet
vertices for both quark\cite{quarkjet} and gluon\cite{gluonjet}
initiated jets at NLO accuracy. \\

In Ref. [\refcite{Chachamis:2012gh,Chachamis:2013hma}] this programme
has been extended to the calculation of the 2-loop gluon Regge
trajectory. This universal function associated with the exchange of a
single reggeized gluon provides a key ingredient in the formulation of
high energy factorization and the resummation of high energy
logarithms of QCD scattering amplitudes.  Multiple reggeized gluon
exchanges appear on the other hand for the high energy description of
the imaginary part of scattering amplitudes and in general for
amplitudes beyond NLL accuracy. While this requires new elements,
which describe in a nutshell the interaction between reggeized gluons,
the gluon Regge trajectory remains an essential building block in the
formulation of high energy resummation also in this more general case.
To be more precise, for the elastic process $p_a + p_b \to p_1 + p_2$
with $s = (p_a + p_b)^2$ and $t = q^2$ with $q = p_a - p_1$ one finds
for amplitudes with gluon quantum numbers in the $t$-channel at LL and
NLL accuracy the following factorized form\footnote{For a pedagogical
  review see Ref.~[\refcite{Fadin:1998sh}].}
\begin{align}
  \label{eq:M8}
 \frac{ \mathcal{M}_{(\bf 8_A)} (s,t)}{ \mathcal{M}^{(0)} (s,t) } & = \Gamma_{a1}(t)  \left[ \left(\frac{-s}{-t} \right)^{\omega(t)}  +   \left(\frac{s}{-t} \right)^{\omega(t)} \right]  \Gamma_{b2}(t), 
\end{align}
where $\mathcal{M}^{(0)}_{(\bf 8_A)}$ is the tree-level amplitude
and the subscript `${\bf 8_A}$' denotes that the allowed $t$-channel
exchange is restricted to the anti-symmetric color octet channel.
The functions $\Gamma_{ij}(t)$ are known as impact factors, describing
the coupling of the reggeized gluons to scattering particles. For the
case of gluon and quarks they have been determined within the
effective action in Ref.~[\refcite{quarkjet,gluonjet}]. The function
$\omega(t)$ which governs the $s$-dependence of the scattering
amplitude is on the other hand the Regge trajectory of the gluon.   
It has been originally
derived in Ref.~[\refcite{Fadin:1996tb,Fadin:1995km}] using $s$-channel
unitarity relations. The result was then subsequently confirmed in
Ref.~[\refcite{Blumlein:1998ib}], clarifying an ambiguity in the  non-infrared divergent contributions of Ref.~[\refcite{Korchemskaya:1996je}]. The
original result was further verified by explicitly evaluating the high
energy limit of 2-loop partonic scattering amplitudes
Ref.~[\refcite{DelDuca:2001gu}]. While the explicit result for the 2-loop
gluon Regge trajectory is by now firmly established, our calculation
provides an important confirmation of its universality: unlike
previous calculations, the effective action defines the Regge
trajectory of the gluon without making any reference to a particular
QCD scattering process.\\

The outline of this contribution is as follows: Sec.~\ref{2} provides
a short introduction to Lipatov's effective action and a list of
necessary Feynman rules, together with a discussion of our
regularization and the employed pole prescription. In
Sec.~\ref{sec:gluontrajectory-effectiveaction} we recall the scheme which we
follow in the derivation of the gluon Regge trajectory and presents
our result.  Sec.~\ref{5} contains our conclusions.

\section{Lipatov's  high energy effective action}
\label{2}

\begin{figure}[htb]
    \label{fig:subfigures}
   \centering
   \parbox{.7cm}{\includegraphics[height = 1.8cm]{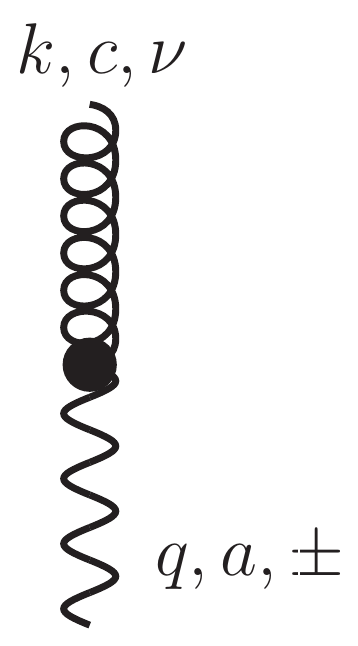}} $=  \displaystyle 
   \begin{array}[h]{ll}
    \\  \\ - i{\bm q}^2 \delta^{a c} (n^\pm)^\nu,  \\ \\  \qquad   k^\pm = 0.
   \end{array}  $ 
 \parbox{1.2cm}{ \includegraphics[height = 1.8cm]{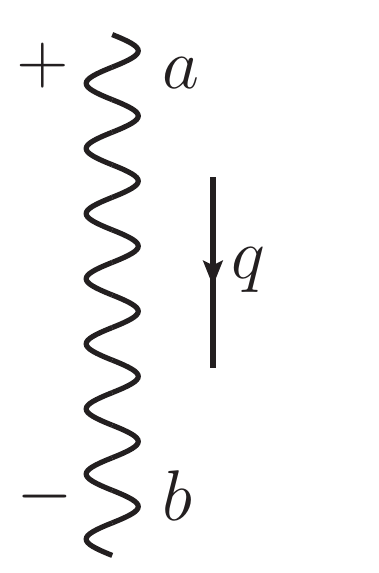}}  $=  \displaystyle    \begin{array}[h]{ll}
    \delta^{ab} \frac{ i/2}{{\bm q}^2} \end{array}$ 
 \parbox{1.7cm}{\includegraphics[height = 1.8cm]{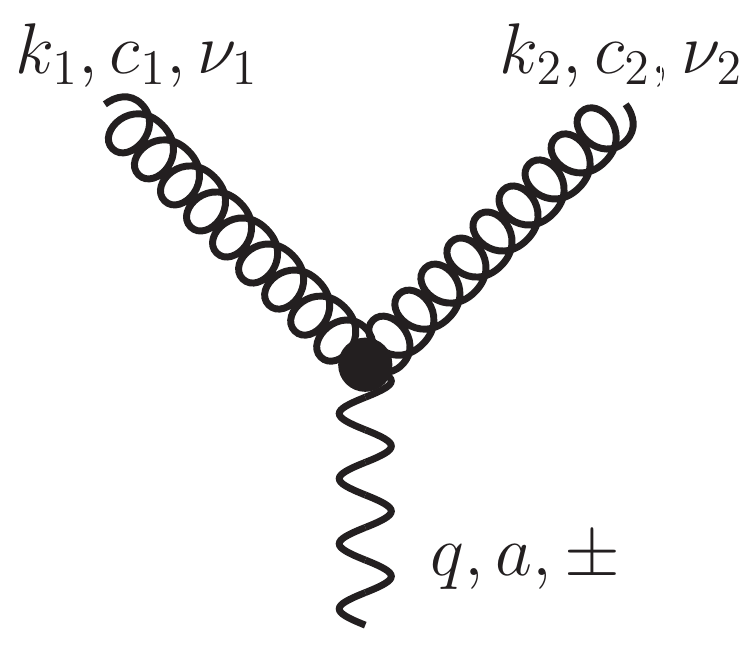}} $ \displaystyle  =  \begin{array}[h]{ll}  \\ \\ g f^{c_1 c_2 a} \frac{{\bm q}^2}{k_1^\pm}   (n^\pm)^{\nu_1} (n^\pm)^{\nu_2},  \\ \\ \quad  k_1^\pm  + k_2^\pm  = 0.
 \end{array}$
 \\
\parbox{4cm}{\center (a)} \parbox{4cm}{\center (b)} \parbox{4cm}{\center (c)}

\vspace{1cm}
  \parbox{2.4cm}{\includegraphics[height = 1.8cm]{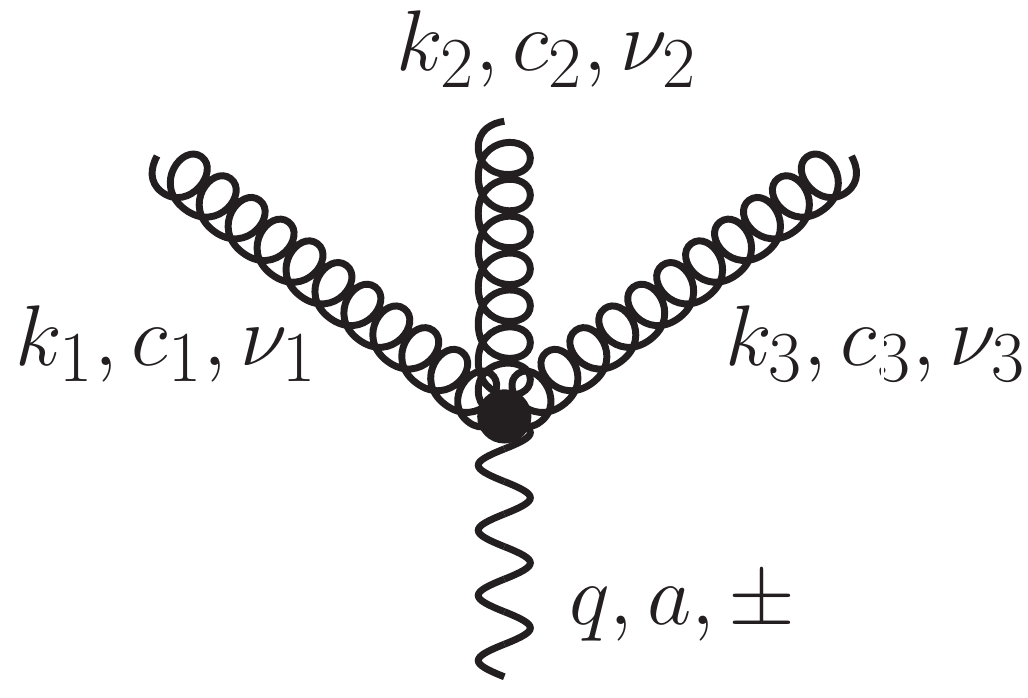}} $ \displaystyle 
   \begin{array}[h]{l}  \displaystyle  \\ \displaystyle= ig^2 {\bm{q}}^2 
\left(\frac{f^{a_3a_2 e} f^{a_1ea}}{k_3^\pm k_1^\pm} 
+
 \frac{f^{a_3a_1 e} f^{a_2ea}}{k_3^\pm k_2^\pm}\right) (n^\pm)^{\nu_1} (n^\pm)^{\nu_2} (n^\pm)^{\nu_3}, \\ \\
\qquad \qquad   k_1^\pm + k_2^\pm + k_3^\pm = 0.
   \end{array}
$ \\ 
\vspace{.3cm}
\parbox{1cm}{(d)}

 \caption{\small Feynman rules for the lowest-order effective vertices of the effective action. Wavy lines denote reggeized fields and curly lines gluons. }
\label{fig:3}
\end{figure}

The effective action\cite{LevSeff} describes interactions which are
restricted to an interval of narrow width ($\eta$) in rapidity
space. Dynamics extending over rapidity separations larger than
$\eta$, is integrated out and taken into account through universal
eikonal factors.  QCD amplitudes are then in the limit of large center
of mass energies constructed through a new degree of freedom ---the
reggeized gluon. The high energy effective action describes the
interaction of this new field with the QCD field content through
adding an induced term $ S_{\text{ind.}}$ to the QCD action
$S_{\text{QCD}}$,
\begin{align}
  \label{eq:effac}
S_{\text{eff}}& = S_{\text{QCD}} +
S_{\text{ind.}}.
\end{align}
The new term $ S_{\text{ind.}}$ describes the coupling of the gluon
field $v_\mu = -it^a v_\mu^a(x)$ to the reggeized gluon field
$A_\pm(x) = - i t^a A_\pm^a (x)$.  High energy factorized amplitudes
reveal strong ordering in plus and minus components of momenta which
is reflected in the following kinematic constraint obeyed by the
reggeized gluon field
\begin{align}
  \label{eq:kinematic}
  \partial_+ A_- (x)& = 0 = \partial_+ A_+(x).
\end{align}
Even though the reggeized gluon field is charged under the QCD gauge
group SU$(N_c)$, it is invariant under local gauge transformations:
$\delta A_\pm = 0$.  Its kinetic term and the gauge invariant coupling
to the QCD gluon field are contained in the induced term,
\begin{align}
\label{eq:1efflagrangian}
  S_{\text{ind.}} = \int \text{d}^4 x \,
\text{tr}\left[\left(W_-[v(x)] - A_-(x) \right)\partial^2_\perp A_+(x)\right]
+ (+) \leftrightarrow (-),
\end{align}
with 
\begin{align}
  \label{eq:funct_expand}
  W_\pm[v(x)] =&
v_\pm(x) \frac{1}{ D_\pm}\partial_\pm
,
&
D_\pm & = \partial_\pm + g v_\pm (x).
\end{align}
Apart from the usual QCD Feynman rules, the Feynman rules of the
effective action comprise the propagator of the reggeized gluon and an
infinite number of so-called induced vertices, which result from the
non-local functional Eq.~\eqref{eq:funct_expand}. Lowest order
vertices and propagators are collected in Fig.~\ref{fig:3} (for
further details we refer to Ref.~[\refcite{Chachamis:2013hma}]).
Loop diagrams of the effective action lead to a new type of
longitudinal divergence which is not present in conventional quantum
corrections to QCD amplitudes. It can be regularized introducing an external parameter $\rho$, evaluated in the limit $\rho \to \infty$, which deforms the light-like vectors $n^\pm$ into
\begin{align}
  \label{eq:deform}
  n^-  & \to n_a = e^{-\rho} n^+ + n^-, \notag \\
  n^+ & \to n_b = n^+ + e^{-\rho} n^-,
\end{align}
without violating the gauge invariance properties of the induced term
Eq.~\eqref{eq:1efflagrangian}.  While it is possible to identify
$\rho$ with a logarithm in $s$ or the rapidity interval spanned by a
certain high energy process, we refrain from such an interpretation
and consider in the following $\rho$ as an external parameter, similar
to the parameter $\epsilon$ in dimensional regularization in $d = 4 +
2 \epsilon$ dimensions.  The evaluation of loop diagrams requires a
prescription to circumvent the light-cone singularities in the induced
vertices shown in Figs.~\ref{fig:3}. A suitable prescription which
preserves the color structure of unregulated vertices has been derived
in Ref.~[\refcite{Hentschinski:2011xg}], by performing (a) the
replacement $D_\pm \to D_\pm \pm \epsilon$ in
Eq.~\eqref{eq:funct_expand} and (b) subsequently projecting out
symmetric color structures, order by order in perturbation theory. The
resulting pole prescription respects then Bose symmetry of the induced
vertices and high energy factorization, for details see
Ref.~[\refcite{Hentschinski:2011xg}].  For the $\mathcal{O}(g)$ vertex, see 
Fig.~\ref{fig:3}.c, this corresponds to replacing the denominator
$1/k_1^\pm$ by a Cauchy principal value. For the $\mathcal{O}(g^2)$
vertex, see Fig.~\ref{fig:3}.c, the combinations of denominators $1/(k_3^\pm
k_1^\pm)$ and $1/(k_3^\pm k_2^\pm)$ are replaced by functions $g_2^\pm
(3,2,1)$ and $g_2^\pm (3,1,2)$ respectively. It is defined as
\begin{align}
  \label{eq:cpv_rep}
g_2^\pm(i,j,m) = 
    \bigg[&  \frac{1}{[k_i^\pm][k_m^\pm]} + \frac{\pi^2}{3}\delta(k_i^\pm)\delta(k_m^\pm) \bigg],
& \frac{1}{[k]} & \equiv \frac{1}{2} \left(\frac{1}{k + i 0} + \frac{1}{k - i0} \right)
\end{align}
with similar functions for the higher order induced vertices.

\section{The gluon Regge trajectory and  the effective action}
\label{sec:gluontrajectory-effectiveaction}

Determination of the gluon trajectory from
the effective action requires
\begin{itemize}
\item to determine  the propagator of the reggeized gluon to the desired order in perturbation theory;
\item to renormalize its rapidity divergences.
\end{itemize}
The gluon Regge trajectory is then identified as the coefficient of
the $\rho$ dependent term in the renormalization factor.  To derive
the 2-loop gluon Regge trajectory we therefore need to determine the
1- and 2-loop self-energies of the reggeized gluon. Within  the
subtraction procedure proposed in Ref.~[\refcite{quarkjet}] this requires
\begin{itemize}
\item
 determination of the self-energy of the reggeized gluon from the effective action, with the reggeized gluon treated as a background field;
\item
subtraction of all  disconnected contributions which contain internal reggeized gluon lines.
\end{itemize}
At 1-loop,
all  diagrams with internal reggeized gluon lines  vanish and no subtraction occurs;  the  contributing diagrams are 
\begin{align}
   \Sigma^{(1)}\left(\rho; \epsilon, \frac{{\bm q}^2}{\mu^2}    \right)   & = 
  \parbox{1cm}{\vspace{0.1cm} \includegraphics[height = 1.8cm]{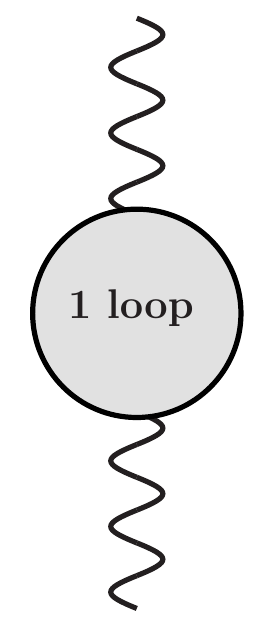}}
= 
    \parbox{.7cm}{\vspace{0.1cm}  \includegraphics[height = 1.8cm]{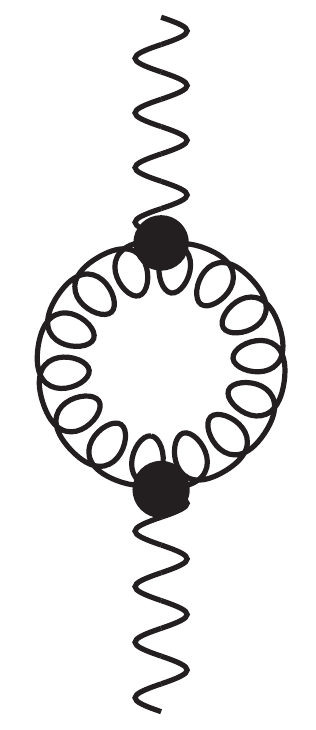}}
  + 
  \parbox{.7cm}{\vspace{0.1cm} \includegraphics[height = 1.8cm]{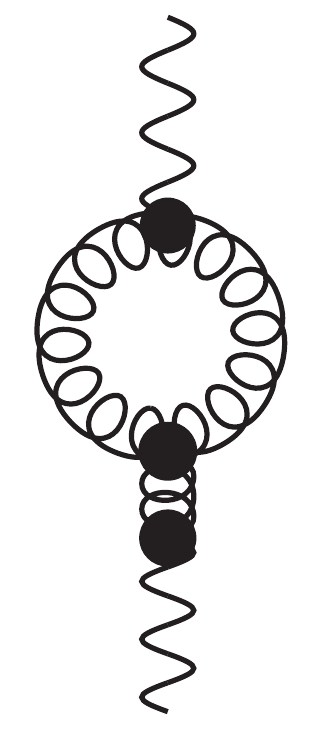}} 
 +
  \parbox{.7cm}{\vspace{0.1cm} \includegraphics[height = 1.8cm]{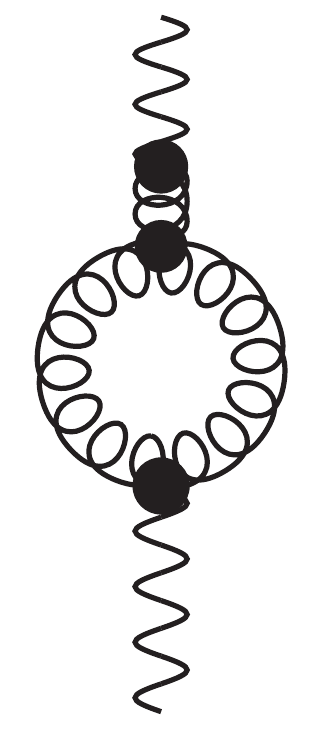}} 
 +
  \parbox{.7cm}{\vspace{0.1cm} \includegraphics[height = 1.8cm]{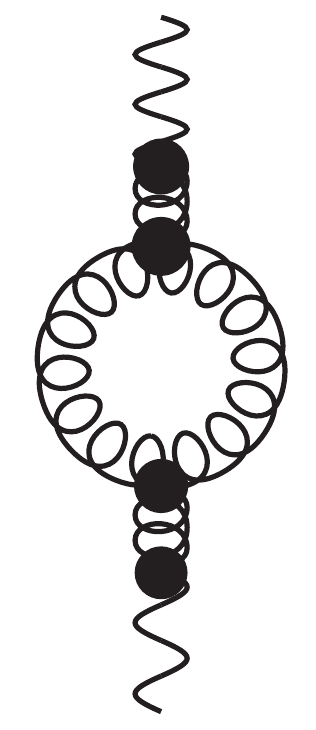}} 
 +
  \parbox{.7cm}{\vspace{0.1cm} \includegraphics[height = 1.8cm]{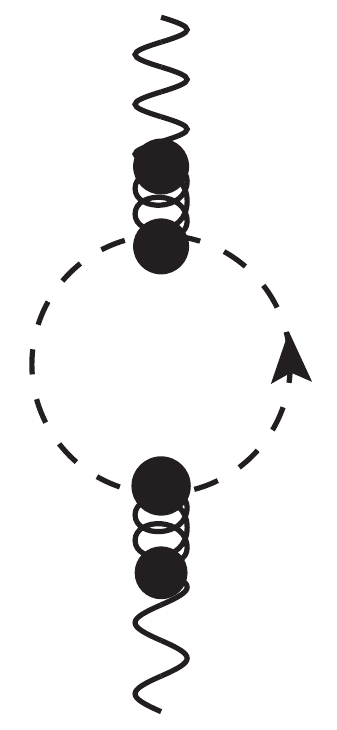}}
+
  \parbox{.7cm}{\vspace{0.1cm} \includegraphics[height = 1.8cm]{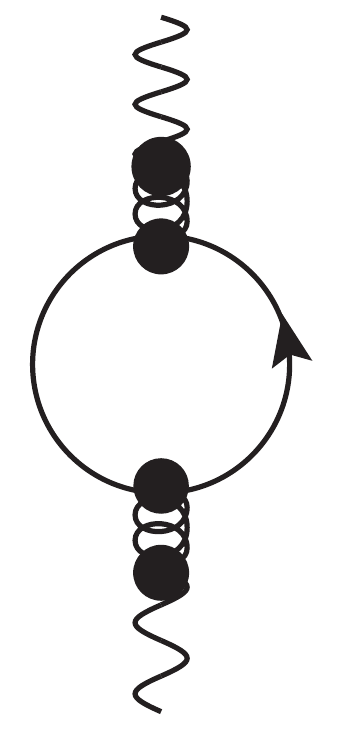}}
\label{eq:self_1loop}
\end{align}
\\
Keeping  for $\rho  \to \infty$ the ${\cal O} (\rho, \rho^2)$  terms and using the notation
\begin{align}
  \label{eq:gbar}
  \bar{g}^2 & =  \frac{g^2 N_c \Gamma(1 - \epsilon)}{(4 \pi)^{2 + \epsilon}}, & c_\Gamma & = \frac{ \Gamma^2(1 + \epsilon)}{\Gamma(1 + 2 \epsilon)},
\end{align}
we have the following result in $d = 4 + 2\epsilon$ dimensions\footnote{In the original result presented in Ref.~[\refcite{quarkjet}] and reproduced in Ref.~[\refcite{Chachamis:2012gh,review}]  a  finite result for the second and third diagram has been erroneously included. This has been corrected in the result presented here.}:
\begin{align}
\label{eq:self_1loop}
   \frac{\Sigma^{(1)}\left(\rho; \epsilon, \frac{{\bm q}^2}{\mu^2}    \right) }{(-2i {\bm q}^2) }     
 &=
\bar{g}^2 c_\Gamma \left(\frac{{\bm q}^2}{\mu^2} \right)^\epsilon  
  \bigg\{  \frac{ i\pi - 2 \rho}{\epsilon}         
- \frac{   5 + 3 \epsilon - \frac{n_f}{N_c} (2 + 2\epsilon)}{(1 + 2 \epsilon)(3 + 2\epsilon)\epsilon} \bigg\}. 
\end{align}
To determine the 2-loop self energy it is on the other hand needed to subtract disconnected diagrams, whereas diagrams with multiple internal reggeized gluons can be shown to yield a zero result. Schematically one has 
\begin{align}
  \label{eq:coeff_2loop}
  \Sigma^{(2)}\left(\rho; \epsilon, \frac{{\bm q}^2}{\mu^2}    \right)   & =   \parbox{2cm}{\center \includegraphics[height = 2.5cm]{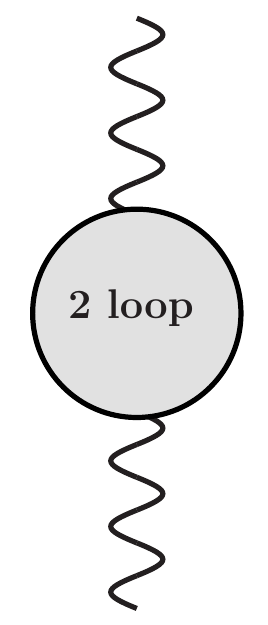}}  
  = 
  \parbox{2cm}{\center \includegraphics[height = 2.5cm]{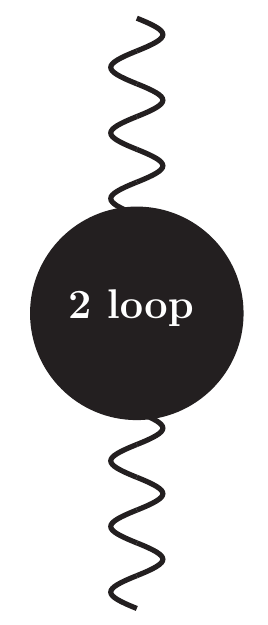}}
  -
  \parbox{2cm}{\center \includegraphics[height = 2.5cm]{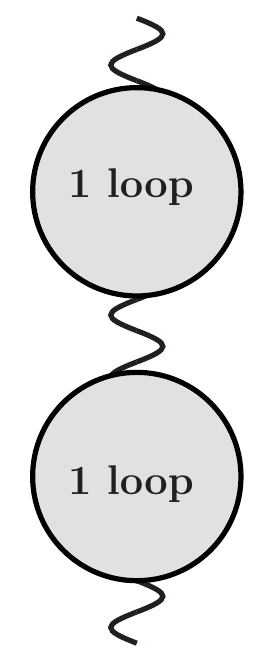}},
\end{align}
where the black blob denotes the unsubtracted 2-loop reggeized gluon
self-energy, which is obtained through the direct application of the
Feynman rules of the effective action, with the reggeized gluon itself
treated as a background field. The emerging set of Feynman integrals
can be reduced to seven master integrals which have been evaluated in
Ref.~[\refcite{Chachamis:2013hma}] using the Mellin-Barnes
representation technique and expansion in $\rho$ using the Mathematica
package \texttt{MBasymptotics.m}\cite{Cza}; the singularities
structure in $\epsilon$ was subsequently resolved using the
Mathematica packages \texttt{MB.m}\cite{Czakon:2005rk} and
\texttt{MBresolve.m}\cite{Smirnov:2009up}.  For details we refer to
Ref.~[\refcite{Chachamis:2012gh,Chachamis:2013hma}]. The result for
$n_f$ flavor reads
\begin{align}
\label{meister}
& \Sigma^{(2)}\left(\rho,\frac{\bm{q}^2}{\mu^2}\right)
= (- 2i{\bm q}^2)\frac{g^4 N_c^2}{(4\pi)^4}
\bigg\{
-\bigg[
\frac{2}{\epsilon^2}+\frac{4(1-\Xi)}{\epsilon}
+ 4(1-\Xi)^2- \frac{\pi^2}{3}
\bigg]\rho^2
\notag \\
&+
\bigg[
 \frac{1}{3\epsilon^2}+\frac{1}{9\epsilon}  +\frac{\pi^2}{3 \epsilon}-\frac{2 \Xi}{3 \epsilon}+\frac{\pi^2(11-12\Xi)}{18}
 +\frac{16}{27} - \frac{2 }{9} \Xi + \frac{2}{3} \Xi^2 -2\zeta(3)\bigg)\bigg]\rho \bigg \} 
\notag \\
&
+ \frac{ n_f}{N_c} \left( \frac{2}{3\epsilon} + \frac{nf (6 - 36 \Xi)}{27 \epsilon} + \frac{32 - 3 \pi^2 - 12 \Xi + 36 \Xi^2}{27}   \right)
+ \mathcal{O}(\epsilon) + \mathcal{O}(\rho^0).&
\end{align}
with $\Xi = 1 - \gamma_E - \ln {\bm q}^2/ (4 \pi \mu^2)$.  The (bare)
two-loop reggeized gluon propagators is then in term of 1- and 2-loop
self energies given by
\begin{align}
  \label{eq:barepropR}
   G \left(\rho; \epsilon, {\bm q}^2, \mu^2   \right)
&=
\frac{i/2}{{\bm q}^2} \left\{ 1 + \frac{i/2}{{\bm q}^2} \Sigma \left(\rho; \epsilon, \frac{{\bm q}^2}{\mu^2}    \right)  + \left[  \frac{i/2}{{\bm q}^2} \Sigma \left(\rho; \epsilon, \frac{{\bm q}^2}{\mu^2}   \right)\right] ^2 + \ldots   \right\},
\end{align}
with
\begin{align}
  \label{eq:sigma}
   \Sigma \left(\rho; \epsilon, \frac{{\bm q}^2}{\mu^2}   \right) & =   \Sigma^{(1)} \left(\rho; \epsilon, \frac{{\bm q}^2}{\mu^2}   \right) +  \Sigma^{(2)} \left(\rho; \epsilon, \frac{{\bm q}^2}{\mu^2}   \right) + \ldots
\end{align}
where the dots indicate higher order terms. Apparently
Eq.~\eqref{eq:self_1loop} is 
divergent in the limit $\rho \to \infty$. In Ref.~[\refcite{quarkjet,gluonjet}]
it has been demonstrated by explicit calculations that these
divergences cancel at one-loop level, for both quark-quark and
gluon-gluon scattering amplitudes, against divergences in the
couplings of the reggeized gluon to external particles. The entire
one-loop amplitude is then found to be free of any high energy
singularity in $\rho$.  High energy factorization then suggests that
such a cancellation holds also beyond one loop. Starting from this assumption, it is possible to define a
renormalized reggeized gluon propagator
\begin{align}
\label{eq:renor}
G^{\rm R}(M^+,M^-;\epsilon,\bm{q}^2,\mu^2)=\frac{G(\rho;\epsilon,\bm{q}^2,\mu^2)}{Z^+\left(\frac{M^+}{\sqrt{\bm{q}^2}},\rho;\epsilon,\frac{\bm{q}^2}{\mu^2}\right)Z^-\left(\frac{M^-}{\sqrt{\bm{q}^2}},\rho;\epsilon,\frac{\bm{q}^2}{\mu^2}\right)},
\end{align}
where the renormalization factors need to cancel against corresponding
renormalization factors associated with the vertex to which the
reggeized gluon couples with `plus' ($Z^+$) and `minus' ($Z^-$)
polarization. For explicit examples we refer the reader to
Ref.~[\refcite{gluonjet,Chachamis:2012gh}].  In their most general form these
renormalization factors are parametrized as
\begin{align}
\label{eq:param}
Z^\pm\left(\frac{M^\pm}{\sqrt{\bm{q}^2}},\rho;\epsilon,\frac{\bm{q}^2}{\mu^2}\right)=\exp\left[\left(\frac{\rho}{2}-\ln\frac{M^\pm}{\sqrt{\bm{q}^2}}\right)\omega\left(\epsilon,\frac{\bm{q}^2}{\mu^2}\right)   +   f^\pm\left(\epsilon,\frac{\bm{q}^2}{\mu^2}\right)\right].
\end{align}
The coefficient of the $\rho$-divergent term defines the gluon Regge trajectory $\omega(\epsilon, {\bm q}^2)$, with the the following perturbative expansion
\begin{align}
  \label{eq:perturexpomega}
  \omega\left(\epsilon, \frac{{\bm q}^2}{\mu^2} \right) &= 
 \omega^{(1)}\left(\epsilon, \frac{{\bm q}^2}{\mu^2} \right)  
+
 \omega^{(2)}\left(\epsilon, \frac{{\bm q}^2}{\mu^2} \right)  + \ldots.
\end{align}
It is  to be determined by the requirement that the renormalized reggeized
gluon propagator must, at each loop order, be free of  $\rho$
divergences.  At one loop one obtains
\begin{align}
  \label{eq:omega1}
  \omega^{(1)}\left(\epsilon, \frac{{\bm q}^2}{\mu^2} \right) & = -\frac{2 \bar{g}^2 \Gamma^2(1 + \epsilon)}{\Gamma(1 + 2 \epsilon)\epsilon } \left(\frac{{\bm q}^2}{\mu^2} \right)^\epsilon.  
\end{align}
The function $f^\pm(\epsilon, {\bm q}^2)$ parametrizes finite
contributions and is, in principle, arbitrary. While symmetry of the
scattering amplitude requires $f^+ = f^- = f$, Regge theory suggests
fixing it in such  that terms which are not enhanced in $\rho$
are entirely transferred from the reggeized gluon propagators to the vertices, to which the reggeized gluon
couples. With the perturbative expansion
\begin{align}
  \label{eq:expansionofF}
  f\left(\epsilon, \frac{{\bm q}^2}{\mu^2} \right) & = f^{(1)}\left(\epsilon, \frac{{\bm q}^2}{\mu^2} \right) + f^{(2)}\left(\epsilon, \frac{{\bm q}^2}{\mu^2} \right) \ldots
\end{align}
we obtain from  Eq.~\eqref{eq:self_1loop}
\begin{align}
  \label{eq:f1loop}
   f^{(1)}\left(\epsilon, \frac{{\bm q}^2}{\mu^2} \right) & =  \frac{ \bar{g}^2 \Gamma^2(1 + \epsilon)}{\Gamma(1 + 2 \epsilon)} \left(\frac{{\bm q}^2}{\mu^2} \right)^\epsilon  
         \frac{(-1)}{(1 + 2 \epsilon)2 \epsilon} \bigg[    \frac{5 + 3\epsilon}{3 + 2 \epsilon} 
-\frac{n_f}{N_c} \left( \frac{2 + 2\epsilon}{3 + 2\epsilon} \right)\bigg]  .
\end{align}
The scales $M^+$ and $M^-$ are arbitrary; their role is analogous to
the renormalization scale in UV renormalization and the factorization
scale in collinear factorization. They are naturally chosen to
be of the order of magnitude of  the corresponding light-cone momenta of scattering
particles to which the reggeized gluon couples. For the 2-loop gluon Regge trajectory  we  obtain the following relation
\begin{align}
  \label{eq:omega2_defined}
 \omega^{(2)}\left(\epsilon, \frac{{\bm q}^2}{\mu^2} \right)  & = 
 \lim_{\rho \to \infty} \frac{1}{\rho} \bigg[ \frac{\Sigma^{(2)}}{(-2i{\bm q}^2)}  + \frac{\rho^2}{2} \left(\omega^{(1)}\right)^2 + 2 \rho   f^{(1)} \omega^{(1)}  \bigg],
\end{align}
where we omitted at the right hand side the dependencies on $\epsilon$
and ${\bm q}^2/\mu^2$ and  expanded
$\Sigma^{(1)}$ in terms of the functions $\omega^{(1)}$ and
$f^{(1)}$. We stress that this is a non-trivial definition and that it
is not clear a priori whether the right hand side even exists due to
the presence of the second term, linear in $\rho$.  Confirmation of
this relation provides therefore an important non-trivial check on the
validity of our formalism. Inserting our result for the 2-loop reggeized gluon self-energy into Eq.~\eqref{eq:omega2_defined} we obtain
\begin{equation}
\omega^{(2)}(\bm{q}^2)=\frac{(\omega^{(1)}(\bm{q}^2))^2}{4}
\left[
\frac{11}{3} - \frac{2 n_f}{3 N_c} +\left(\frac{\pi^2}{3}-\frac{67}{9}\right)\epsilon+\left(\frac{404}{27}-2\zeta(3)\right)\epsilon^2
\right].
\end{equation}
which is in complete agreement with the results in the literature\cite{Fadin:1996tb}.

\section{Conclusions}\label{5}

In this contribution the central steps of the derivation of the
two-loop gluon Regge trajectory from Lipatov's effective action have
been presented. They consist of determination of the reggeized gluon
propagator to the desired order in perturbation theory and subsequent
renormalization of high energy divergences. The coefficient of the
high energy divergence is then identified as the gluon Regge
trajectory. Determination of the reggeized gluon propagator requires
on the other hand evaluation of corresponding effective action
diagrams, combined with a subtraction mechanism.  The final result is
in precise agreement with earlier results present in the literature
and thus serves to further validate the effective action and our
proposed computational framework.

\section*{Acknowledgements}

We thank J. Bartels, V. Fadin and L. Lipatov for constant support for
many years.  We acknowledge partial support by the Research Executive
Agency (REA) of the European Union under the Grant Agreement number
PITN-GA-2010-264564 (LHCPhenoNet), the Comunidad de Madrid through
Proyecto HEPHACOS ESP-1473, by MICINN (FPA2010-17747), by the Spanish
Government and EU ERDF funds (grants FPA2007-60323, FPA2011-23778 and
CSD2007- 00042 Consolider Project CPAN) and by GV
(PROMETEUII/2013/007).  G.C. acknowledges support from Marie Curie
Actions (PIEF-GA-2011-298582). M.H.  acknowledges support from the
U.S. Department of Energy under contract number DE-AC02-98CH10886 and
a ``BNL Laboratory Directed Research and Development'' grant (LDRD
12-034).

\end{document}